\documentclass[aps,pra,twocolumn,superscriptaddress]{revtex4}
\usepackage{times}
\usepackage{latexsym}
\usepackage{graphicx}
\usepackage{amsmath,amssymb}
\usepackage{verbatim,bbm}
\usepackage{amsmath}
\usepackage{amsfonts}
\usepackage{amssymb}
\begin{document}
\title{Channel Simulation in Quantum Metrology}
\author{Riccardo Laurenza}
\affiliation{Computer Science and York Centre for Quantum Technologies, University of York,
York YO10 5GH, UK}
\author{Cosmo Lupo}
\affiliation{Computer Science and York Centre for Quantum Technologies, University of York,
York YO10 5GH, UK}
\author{Gaetana Spedalieri}
\affiliation{Computer Science and York Centre for Quantum Technologies, University of York,
York YO10 5GH, UK}
\affiliation{Research Laboratory of Electronics, Massachusetts Institute of Technology,
Cambridge, Massachusetts 02139, USA}
\author{Samuel L. Braunstein}
\affiliation{Computer Science and York Centre for Quantum Technologies, University of York,
York YO10 5GH, UK}
\author{Stefano Pirandola}
\email{stefano.pirandola@york.ac.uk}
\affiliation{Computer Science and York Centre for Quantum Technologies, University of York,
York YO10 5GH, UK}

\begin{abstract}
In this review we discuss how channel simulation can be used to simplify the
most general protocols of quantum parameter estimation, where unlimited
entanglement and adaptive joint operations may be employed. Whenever the
unknown parameter encoded in a quantum channel is completely transferred in an
environmental program state simulating the channel, the optimal adaptive
estimation cannot beat the standard quantum limit. In this setting, we
elucidate the crucial role of quantum teleportation as a primitive operation
which allows one to completely reduce adaptive protocols over suitable
teleportation-covariant channels and derive matching upper and lower bounds
for parameter estimation. For these channels, we may express the quantum
Cram\'{e}r Rao bound directly in terms of their Choi matrices. Our review
considers both discrete- and continuous-variable systems, also presenting some
new results for bosonic Gaussian channels using an alternative sub-optimal
simulation. It is an open problem to design simulations for quantum channels
that achieve the Heisenberg limit.

\end{abstract}
\maketitle

\section{Introduction}

Quantum technologies exploit quantum
information~\cite{NiCh,RMPwee,RMPcosmo,Ulrikreview} to develop new powerful
devices that aim at solving long-standing problems as well as providing
completely novel applications. This is happening in many areas, including
quantum communication~\cite{Chack,BK,teleREVIEW,Kimble,HybridINTERNET,Vaid},
secret key
distribution~\cite{crypt1,crypt2,crypt3,crypt4,crypt5,crypt6,crypt7,crypt8,crypt9,crypt10,crypt11}%
, sensing (e.g., quantum illumination~\cite{Qill0,Qill1,Qill2,Qill3}), imaging
(e.g., optical resolution~\cite{Makei,Lupo16,Nair}), and
metrology~\cite{BC,met2,met3,met4,Paris,met6,BraunRMP,Lloyd}. The latter area
is particularly active and promising in terms of practical applications.
Quantum metrology~\cite{BC}, also known as quantum parameter estimation, deals
with the estimation of unknown classical parameters which are encoded in
quantum states or quantum transformations, i.e., quantum
channels~\cite{RMPcosmo}. Here we are interested in the latter scenario of
quantum channel estimation. In this setting, we review techniques of channel
simulation~\cite{Gatearray,Qsim0,Matsumoto,tools,metro,nostro} that allow one
to simplify the structure of the most general protocols of quantum parameter
estimation to a much simpler and treatable version.\

To clarify the context, let us formulate the general problem. Suppose that we
are given a black-box implementing a quantum channel $\mathcal{E}_{\theta}$
with an unknown classical parameter $\theta$ with uniform prior. We are then
asked to probe the box $n$ times with the aim of retrieving the best value of
$\theta$. Statistically, this means to generate an estimator $\tilde{\theta}$
of $\theta$ such that its error variance $\delta\theta^{2}=\langle
(\tilde{\theta}-\theta)^{2}\rangle$ is the minimal possible (here the average
is assumed over the $n$ probings of the box). It is clear that $\delta
\theta^{2}$, or the standard deviation $\delta\theta$, is expected to decrease
as a function of $n$. Therefore an important crucial question to answer is the
following: What is the optimal scaling in $n$?

For certain channels the optimal scaling is $\delta\theta\sim n^{-1/2}$, known
as the \textquotedblleft standard quantum limit\textquotedblright\ (SQL)
because it is also what you would aspect with in a completely classical
setting. Remarkably, this limit can be beaten for other channels, so that they
display a fully quantum behaviour. In fact, it is known that the optimal
scaling that is reachable in the quantum setting is $\delta\theta\sim n^{-1}$,
also called the \textquotedblleft Heisenberg limit\textquotedblright%
\ (HL)~\cite{Lloyd}. In order to understand if a channel $\mathcal{E}_{\theta
}$ is limited to the SQL or not, it is essential to adopt the most general
quantum protocols of parameter estimation that are allowed by quantum
mechanics. These protocols involve the use of unlimited entanglement and are
inevitably adaptive, i.e., may involve the use of joint quantum operations
where the inputs to the box are optimised as a result of all the previous
rounds~\cite{metro,nostro,RafalPRX,Preskill}. It is clear that their study is
extremely difficult and require some techniques that may reduce their
complexity. In this respect, channel simulation is certainly one of the most
powerful tools.

Here we review the most important results for channel simulation in quantum
metrology, plus we present some new bounds. We start with the discussion of
programmable channels~\cite{Gatearray,Qsim0}, which are those channels
$\mathcal{E}$ that can be simulated by a program state $\pi_{\mathcal{E}}$ and
some joint (trace-preserving) quantum operation or \textquotedblleft
simulator\textquotedblright\ $\mathcal{S}$ applied to the input state $\rho$
and the program $\pi_{\mathcal{E}}$, so that $\mathcal{E}(\rho)=\mathcal{S}%
(\rho\otimes\pi_{\mathcal{E}})$. When a parameter $\theta$ labels the channel
$\mathcal{E}_{\theta}$, it may happen that the previous simulator
$\mathcal{S}$ remains \textquotedblleft universal\textquotedblright, i.e.,
independent on $\theta$, while the program state completely absorbs the label,
i.e., it becomes $\pi_{\mathcal{E}_{\theta}}$. If this is the case, one may
re-organise an adaptive protocol in a block version and show that the SQL is
an upper bound that cannot be beaten~\cite{metro,nostro}.

Recently, Ref.~\cite{nostro} adopted a simple criterion to identify these
channels at any dimension (finite or infinite). Whenever a quantum channel is
teleportation covariant~\cite{PLOB}, i.e., suitably commuting with
teleportation unitaries, it can be simulated by teleporting over its Choi
matrix, i.e., the simulator $\mathcal{S}$ is teleportation and the program
state $\pi_{\mathcal{E}_{\theta}}$ is the channel's Choi matrix~\cite{nostro}.
Thus for these channels, we have a precise and simple design for their
simulation. Furthermore, this design allows one to show that the SQL\ is
asymptotically achievable with a prefactor which is completely determined by
the Choi matrix of the channel. Thus, Ref.~\cite{nostro} showed that
teleportation-covariance implies the SQL, elucidating how teleportation gives
a \textit{no-go} for Heisenberg scaling.

The methodology of Ref.~\cite{nostro} applies to quantum channels of any
dimension. As we will explain, the teleportation simulation of bosonic
channels~\cite{RMPwee} needs a careful treatment due to the fact that both the
ideal maximally-entangled state and the ideal Bell detection require infinite
energy in the setting of continuous-variable systems. Therefore, suitable
limits and truncations of the Hilbert spaces need to be considered to avoid
divergences~\cite{nostro,PLOB}. Besides specifying these aspects, we also
exploit a different sub-optimal simulation of these channels, where asymptotic
maximally-entangled states are not needed, following a recent
approach~\cite{finite}.

The paper is structured as follows. In Sec.~\ref{Sec1}, we review strategies
of quantum parameter estimation giving the main definitions. In
Sec.~\ref{Sec2}\ we discuss the simulation of programmable channels and their
restriction to the SQL. We also discuss potential extensions of this
simulation. Then, in Sec.~\ref{Sec3}, we introduce the specific teleportation
design, valid for teleportation-covariant channels, and the teleportation
stretching of the parameter estimation protocol. We extend these tools to
continuous variable systems and bosonic channels in Sec.~\ref{Sec4}. Then, in
Sec.~\ref{Sec5}, we present some novel bounds based on sub-optimal simulations
of Gaussian channels. Finally, Sec.~\ref{Sec6} is for conclusions.

\section{Protocols of quantum parameter estimation\label{Sec1}}

As already mentioned in the introduction, consider the scenario where we are
given a black-box whose input-output physical transformation can be modelled
as a quantum channel $\mathcal{E}_{\theta}$ encoding an unknown parameter
$\theta$ with uniform prior distribution (i.e., completely random). The task
is to infer $\theta$ with an optimal estimator $\tilde{\theta}$, i.e., with
minimal error variance $\delta\theta^{2}$. It is clear that the performance
will depend on the specific probing strategy which is adopted. The most basic
operations to be done are: (1) Preparing a suitable input state to probe the
channel; and (2) detecting the output of the channel by means of a suitable
measurement or positive operator-valued measure (POVM). These elementary
operations are the only ones that are exploited in block protocols of
parameter estimation, which may be \textquotedblleft direct\textquotedblright%
\ and \textquotedblleft assisted\textquotedblright.

\begin{figure}[ptb]
\vspace{+0.1cm} \centering
\includegraphics[width=0.25\textwidth]{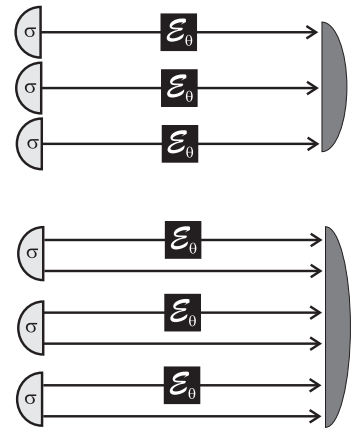}\caption{Block protocols for
quantum parameter estimation, i.e., the direct protocol (top) and the assisted
protocol (bottom). In these protocols, $n$ instances of the quantum channel
$\mathcal{E}_{\theta}$ are identically and independently probed with the same
input state $\sigma$. The resulting output state is a tensor product which is
subject to an optimal POVM, whose output is post-processed into an (unbiased)
estimator $\tilde{\theta}$ of $\theta$.}%
\label{Fig:tp}%
\end{figure}

A direct protocol is shown in Fig.~\ref{Fig:tp}(top). For each of the $n$
probings of the channel $\mathcal{E}_{\theta}$, we prepare the same input
state $\sigma$, so that the total output is a tensor product state
$\rho_{\theta}^{\otimes n}=\mathcal{E}_{\theta}(\sigma)^{\otimes n}$, which is
then detected by a joint POVM. An assisted protocol is shown in
Fig.~\ref{Fig:tp}(bottom). In each probing of the channel we use a joint state
$\sigma$ of the input system and an ancillary system. Therefore, the total
output state has a slightly different tensor product form $\rho_{\theta
}^{\otimes n}=[(\mathcal{E}_{\theta}\otimes I)(\sigma)]^{\otimes n}$. This
state is then jointly measured. It is clear that an assisted protocol is a
direct protocol over the extended channel $\mathcal{E}_{\theta}\otimes I$.

The most general protocol of quantum parameter estimation involves additional
ingredients. Each probing of the channel may in fact be interleaved with joint
quantum operations. In this way, unlimited entanglement may be distributed
between input and output, and feedback may also be used to adaptively optimise
the inputs of the next transmissions~\cite{metro,nostro}. We may think to have
a quantum register prepared in some fundamental initial state $\sigma$. After
a first joint operation, one system is picked from this register and
transmitted through the channel. The output becomes again part of the
register, which is collectively subject to another joint quantum operation.
Then, there is the second probing by transmitting another system of the
register through the channel and so on. After $n$ such adaptive probings, we
have an output state $\rho_{\theta}^{n}$ which is subject to a joint POVM.
Note that we may assume that the adaptive quantum operations are
trace-preserving, because any non-trace preserving process can always be
delayed and included in the final POVM by the principle of deferred
measurement~\cite{NiCh}.\begin{figure}[ptb]
\centering
\includegraphics[width=0.32\textwidth]{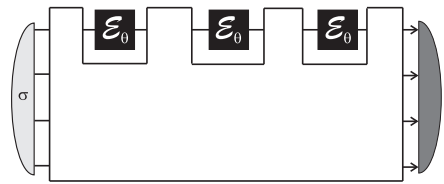}\caption{The most general
(adaptive) estimation protocol can be represented as a quantum comb, i.e., a
quantum circuit board with $n$ slots to plug $n$ instances of the channel
$\mathcal{E}_{\theta}$ in. The initial state of the quantum comb is denoted as
$\sigma$ and the output state as $\rho_{\theta}^{n}$. The output state is
finally detected by a joint POVM whose outcome is classically post-processed
to estimate $\theta$.}%
\label{Fig:comb}%
\end{figure}

An equivalent way to present this adaptive protocol is by resorting to the
model of quantum comb~\cite{qcomb}, as shown in Fig.~\ref{Fig:comb}. Indeed, a
quantum comb represents a quantum circuit board with $n$ slots to plug in $n$
instances of the quantum channel. The internal structure of the quantum comb
is completely generic and includes any possible quantum gate. The initial
state $\sigma$ of the quantum comb is transformed into an output state
$\rho_{\theta}^{n}$ after the action of the comb and the channel. The final
state of the comb is then detected by a joint POVM, whose outcome is processed
into an estimator $\tilde{\theta}$. Note that this strategy includes the
previous block protocols as particular cases. It also includes the so-called
\textquotedblleft sequential\textquotedblright\ protocols~\cite{metro}, where
a state is transmitted through the entire sequence of $n$ channels before detection.

Suppose that we implement an optimal adaptive protocol, i.e., we implicitly
optimise over all possible quantum combs and all possible joint POVMs. The
ultimate lower bound for the error variance of any unbiased estimator is the
quantum Cram\'{e}r-Rao bound (QCRB)%
\begin{equation}
\delta\theta^{2}\geq\frac{1}{\mathrm{QFI}(\rho_{\theta}^{n})},
\end{equation}
where $\mathrm{QFI}$ denotes the quantum Fisher information~\cite{BC}
\begin{equation}
\mathrm{QFI}(\rho_{\theta}^{n})=\mathrm{Tr}\left(  \mathcal{L}_{\theta}%
^{2}\rho_{\theta}^{n}\right)  ,
\end{equation}
and $\mathcal{L}_{\theta}$ is the symmetric logarithmic derivative (SLD).

Assuming that the output has spectral decomposition
\begin{equation}
\rho_{\theta}^{n}=\sum_{j}\lambda_{j}|e_{j}\rangle\langle e_{j}|,
\end{equation}
the expression of the SLD is given by~\cite{BC,Paris}
\begin{equation}
\mathcal{L}_{\theta}=\sum_{j,k:\lambda_{j}+\lambda_{k}>0}\frac{2}{\lambda
_{j}+\lambda_{k}}\,\langle e_{j}|\frac{d\rho_{\theta}^{n}}{d\theta}%
|e_{k}\rangle\,|e_{j}\rangle\langle e_{k}|.
\end{equation}
Alternatively, we may express the QFI as~\cite{BC}
\begin{equation}
\mathrm{QFI}(\rho_{\theta}^{n})=\frac{8[1-F(\rho_{\theta}^{n},\rho
_{\theta+d\theta}^{n})]}{d\theta^{2}},
\end{equation}
where $F(\rho,\sigma):=\mathrm{Tr}\sqrt{\sqrt{\sigma}\rho\sqrt{\sigma}}$ is
the quantum fidelity~\cite{Fid1,Fid2}, which is known to have closed
analytical forms, e.g., for two arbitrary Gaussian states~\cite{Banchi}.

It is important to recall two fundamental properties of the QFI. The first one
is its additivity over tensor products. Given any two parametrised states
$\gamma_{\theta}$ and $\gamma_{\theta}^{\prime}$, we may write
\begin{equation}
\mathrm{QFI}(\gamma_{\theta}\otimes\gamma_{\theta}^{\prime})=\mathrm{QFI}%
(\gamma_{\theta})+\mathrm{QFI}(\gamma_{\theta}^{\prime})~.
\end{equation}
The second is its monotonicity under completely-positive and trace preserving
(CPTP) maps, i.e., quantum channels. Given a quantum channel $\Lambda$, we may
write
\begin{equation}
\mathrm{QFI}[\Lambda(\gamma_{\theta})]\leq\mathrm{QFI}(\gamma_{\theta})~.
\end{equation}

Note that, because the output of a block protocol (direct or assisted) is a
tensor product state $\rho_{\theta}^{\otimes n}$ and the additivity of the QFI
implies $\mathrm{QFI}(\rho_{\theta}^{\otimes n})=n\mathrm{QFI}(\rho_{\theta}%
)$, we have that the QCRB associated with this type of protocol becomes
\begin{equation}
\delta\theta^{2}\geq\frac{1}{n\mathrm{QFI}(\rho_{\theta})},
\end{equation}
so that it scales according to the SQL.

By contrast, the output $\rho_{\theta}^{n}$ of an adaptive protocol is not
necessarily a product state. For this reason, the error variance may behave
differently and potentially beat the SQL. Indeed, it is known that
$\delta\theta^{2}$ may scale according to the HL, for instance in the
estimation of the phase in a unitary transformation~\cite{Lloyd}. However, the
possibility to express the output state $\rho_{\theta}^{n}$ as a quantum
channel applied to a tensor product, i.e., $\rho_{\theta}^{n}=\Lambda
(\gamma_{\theta}^{\otimes n})$, automatically reduces the performance of the
protocol back to the SQL, because of the monotonicity and additivity of the
QFI. In fact, we may write $\mathrm{QFI}(\rho_{\theta}^{n})\leq\mathrm{QFI}%
(\gamma_{\theta}^{\otimes n})=n\mathrm{QFI}(\gamma_{\theta})$. In the
following section, we discuss the conditions for this reduction.

\section{Programmable channels and protocol reduction\label{Sec2}}

Here we discuss how the most general adaptive protocol for quantum parameter
estimation (as the comb in Fig.~\ref{Fig:comb}) can be reduced to a block
protocol when implemented over programmable channels. This implies that
quantum metrology with programmable channels is bounded to the SQL.

The original idea of programmability was introduced by Nielsen and
Chuang~\cite{Gatearray} in the context of quantum computation. These authors
introduced a model of programmable quantum gate array (PQGA) for the
simulation of an arbitrary quantum channel by using a universal unitary and a
program state. Assuming finite resources (e.g., a finite number of systems for
the program state), the simulation can only be probabilistic. Alternatively,
an arbitrary quantum channel can be simulated if we are allowed to use an
infinite number of systems (note that this is exactly the limit which needs to
be taken in the equivalent formulation of port-based
teleportation~\cite{PBT,PBT1,PBT2,PBT3} if one wants to achieve perfect fidelity).

Later in 2008, Ref.~\cite{Qsim0}\ considered a variant of the PQGA
where the simulation is deterministic but can only be applied to a
subset of channels, called \textquotedblleft
programmable\textquotedblright\ channels. This tool was used in
the context of quantum metrology but not immediately applied to
adaptive protocols. See also Ref.~\cite{Matsumoto}. It was later
called \textquotedblleft quantum simulation\textquotedblright\ in
Ref.~\cite{tools}. The first application to simplify adaptive
protocols was presented in Ref.~\cite{metro}\ in the context of
discrete-variable channels. Later, Ref.~\cite{nostro} considered
programmable channels in the context of both discrete- and
continuous-variable channels, also identifying the crucial
connection with quantum teleportation that we will describe
later.\begin{figure}[ptb] \centering \vspace{+0.2cm}
\includegraphics[width=0.27\textwidth]{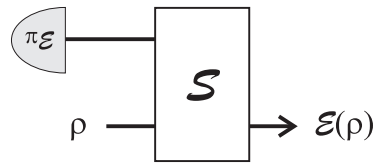} \vspace{-0.1cm}
\caption{A programmable channel admits a simulation of the form
$\mathcal{E}(\rho)=\mathcal{S}(\rho\otimes\pi_{\mathcal{E}})$
where $\mathcal{S}\ $is a simulation channel and
$\pi_{\mathcal{E}}$ a program state. Channels are co-programmable
when they have the same $\mathcal{S}$, but
generally different program states.}%
\label{Fig:program}%
\end{figure}

A quantum channel $\mathcal{E}$ is called programmable if there is a
\textquotedblleft simulator\textquotedblright\ $\mathcal{S}$ (another quantum
channel) and a program state $\pi_{\mathcal{E}}$, such that
\begin{equation}
\mathcal{E}(\rho)=\mathcal{S}(\rho\otimes\pi_{\mathcal{E}}). \label{progdd}%
\end{equation}
This is also shown in Fig.~\ref{Fig:program}. Without loss of generality, the
channel simulator can always be dilated into a unitary $U$ up to introducing
extra degrees of freedom in the program state. Then we also say that an
ensemble of channels $\Omega$ is \textquotedblleft
co-programmable\textquotedblright\ if the simulator $\mathcal{S}$ is universal
over $\Omega$. In other words, for any $\mathcal{E}\in\Omega$, we may write
Eq.~(\ref{progdd}) with exactly the same $\mathcal{S}$ but generally-different
program states $\pi_{\mathcal{E}}$.

\begin{figure}[ptb]
\centering \vspace{+0.2cm}
\includegraphics[width=0.42\textwidth]{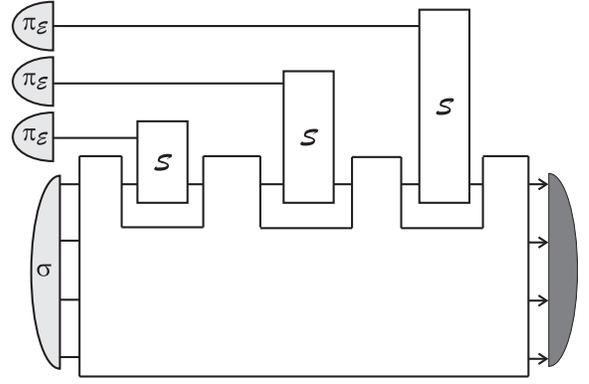}\caption{Stretching of a
quantum comb. First, suppose we have a quantum comb whose slots are filled by
a programmable channel $\mathcal{E}$. Using the simulation of
Fig.~\ref{Fig:program}, we may replace $n$ instances of the quantum channel
$\mathcal{E}$ with a collection of $n$ programme states $\pi_{\mathcal{E}}$.
The corresponding simulators $S$ can be included in the operations of the
quantum comb. In this way, we may collapse the quantum comb (including its
initial state $\sigma$) and the simulators into a single quantum channel
$\Lambda$ applied to the programme states, so that the output of the comb
$\rho^{n}$ can be decomposed as $\Lambda(\pi_{\mathcal{E}}^{\otimes n})$. In
the setting of adaptive parameter estimation, the slots of the comb are filled
by a quantum channel $\mathcal{E}_{\theta}$ encoding the unknown classical
parameter $\theta$. Assuming that $\mathcal{E}_{\theta}$ spans a family of
co-programmable channels, then we may repeat the procedure, and decompose the
output state $\rho_{\theta}^{n}$ as $\Lambda(\pi_{\mathcal{E}_{\theta}%
}^{\otimes n})$. }%
\label{Fig:stretch}%
\end{figure}

Let us now apply these notions to parameter estimation. Assume that the
parametrised quantum channel $\mathcal{E}_{\theta}$ spans a family of
co-programmable channels, so that we may write the simulation
\begin{equation}
\mathcal{E}_{\theta}(\rho)=\mathcal{S}(\rho\otimes\pi_{\mathcal{E}_{\theta}%
}),~~\text{for any~}\theta\text{.} \label{simttt}%
\end{equation}
We can then simplify any adaptive protocol over $n$ uses of this channel. In
fact, we may replace each instance of the channel with its simulation of
Fig.~\ref{Fig:program}, so that the quantum comb in Fig.~\ref{Fig:comb} can be
re-organised in the form depicted in Fig.~\ref{Fig:stretch}. The idea is to
replace each use of the channel $\mathcal{E}_{\theta}$ with its program state
$\pi_{\mathcal{E}_{\theta}}$, and then to \textquotedblleft
stretch\textquotedblright\ all the program states back in time, while
collapsing the simulators $\mathcal{S}$ and the quantum comb (including its
initial state $\sigma$) into a single quantum channel $\Lambda$. In this way,
the output of the comb can be decomposed as
\begin{equation}
\rho_{\theta}^{n}=\Lambda(\pi_{\mathcal{E}_{\theta}}^{\otimes n})~.
\label{decompll}%
\end{equation}

Note that the latter decomposition reduces the adaptive protocol into a block
protocol up to an overall quantum channel $\Lambda$. Because of the properties
of the QFI, we know that this is sufficient to restrict the performance of the
protocol to the SQL. In fact, using monotonicity and additivity of \ the QFI,
we may write
\begin{equation}
\mathrm{QFI}(\rho_{\theta}^{n})=\mathrm{QFI}[\Lambda(\pi_{\mathcal{E}_{\theta
}}^{\otimes n})]\leq\mathrm{QFI}(\pi_{\mathcal{E}_{\theta}}^{\otimes
n})=n\mathrm{QFI}(\pi_{\mathcal{E}_{\theta}}). \label{ub4}%
\end{equation}
We have thus obtained that for the estimation of a parameter $\theta$ encoded
in a programmable channel $\mathcal{E}_{\theta}$, the QCRB must satisfy the
condition $\delta\theta^{2}\geq\lbrack n\mathrm{QFI}(\pi_{\mathcal{E}_{\theta
}})]^{-1}$. Furthermore, note that this bound is not necessarily achievable.
It would be achievable if the program state $\pi_{\mathcal{E}_{\theta}}$ could
be generated by sending some input state through the channel. For instance,
this would be the case if $\pi_{\mathcal{E}_{\theta}}$ were the Choi matrix of
the channel, an extra property which is guaranteed if the channel is
teleportation-covariant, as explained in the next section.

Before proceeding, we may ask how the channel simulation should be modified in
order to cover channels that beat the SQL. One potential idea is to weaken the
notion of co-programmability to involve multi-copy program states. For
instance, suppose that a quantum channel $\mathcal{E}_{\theta}$ cannot be
simulated as in Eq.~(\ref{simttt}) but as%
\begin{equation}
\mathcal{E}_{\theta}(\rho)=\mathcal{S}(\rho\otimes\pi_{\mathcal{E}_{\theta}%
}^{\otimes m}),~~\text{for any~}\theta\text{ and some }m\text{.} \label{conhh}%
\end{equation}
This leads to the stretching $\rho_{\theta}^{n}=\Lambda(\pi_{\mathcal{E}%
_{\theta}}^{\otimes mn})$ and therefore to
\begin{equation}
\mathrm{QFI}(\rho_{\theta}^{n})\leq mn\mathrm{QFI}(\pi_{\mathcal{E}_{\theta}%
}).
\end{equation}
We know that the HL $\delta\theta^{2}\gtrsim n^{-2}$ cannot be beaten so that
we must have $m\leq n$. To get the HL it is sufficient that the condition in
Eq.~(\ref{conhh}) holds asymptotically, i.e., in trace norm limit $\delta
_{m}:=||\mathcal{E}_{\theta}(\rho)-\mathcal{S}(\rho\otimes\pi_{\mathcal{E}%
_{\theta}}^{\otimes m})||\overset{m}{\rightarrow}0$. Then we may take this
limit jointly with the limit in $n$ for the scaling.

\section{Teleportation simulation\label{Sec3}}

Teleportation simulation has been used in the past to reduce protocols of
quantum communication into entanglement
distillation~\cite{BDSW,Wolfnotes,MHthesis,Leung} and, more recently, to
completely simplify protocols of private communication from adaptive to block
forms~\cite{PLOB}, establishing the ultimate limits of QKD\ in point-to-point
lossy communications~\cite{PLOB} and also multi-point~\cite{multipoint} and
repeater-assisted scenarios~\cite{netPIRS}. More recently, Ref.~\cite{nostro}
extended the technique to quantum metrology and quantum channel discrimination.

Let us start with discrete-variable systems and, in particular, qubits
(arguments can be easily generalised to any finite dimension). We first recall
the basic ingredients of teleportation and then we discuss how these can be
modified to implement a tool of channel simulation. The standard qubit
teleportation protocol~\cite{Chack,PirManci} can be broken down in three steps:

\begin{description}
\item[(1) Resource.] A maximally-entangled state $|\Phi_{+}\rangle
=(|00\rangle+|11\rangle)/\sqrt{2}$ is prepared for qubits $A$ (Alice) and $B$ (Bob).

\item[(2) Bell detection.] Alice performs a Bell detection on qubit $A$ and an
input qubit $a$ (in an arbitrary state $\rho$). Recall that the Bell detection
has four outcomes $\alpha\in\{0,1,2,3\}$ with POVM elements $|\Phi_{\alpha
}\rangle\langle\Phi_{\alpha}|$ where $|\Phi_{\alpha}\rangle=(I\otimes
\sigma_{\alpha})|\Phi_{+}\rangle$ and $\sigma_{\alpha}$ are the four Pauli
unitaries~\cite{NiCh}
\begin{align}
&  \sigma_{0}=I:=\left(
\begin{array}
[c]{cc}%
1 & 0\\
0 & 1
\end{array}
\right)  ,\,\,\sigma_{1}=X:=\left(
\begin{array}
[c]{cc}%
0 & 1\\
1 & 0
\end{array}
\right)  ,\\
&  \sigma_{2}=Y:=\left(
\begin{array}
[c]{cc}%
0 & i\\
-i & 0
\end{array}
\right)  ,\,\,\sigma_{3}=Z:=\left(
\begin{array}
[c]{cc}%
1 & 0\\
0 & -1
\end{array}
\right)  .
\end{align}

\item[(3) Pauli corrections.] Finally, depending on the output of the Bell
measurement $\alpha$, the conditional Pauli unitary $\sigma_{\alpha}^{-1}$ is
applied on the qubit $B$, retrieving the input state $\rho$.
\end{description}

The standard teleportation protocol simulates the identity channel. A
modification of the protocol is to employ a resource state which is not
maximally-entangled but an arbitrary bipartite state. In this way
teleportation implements not the identity but simulates a noisy channel from
the input qubit $a$ to the output qubit $B$. Suppose that we choose the
resource state to be the Choi matrix of a quantum channel $\mathcal{E}$,
i.e.,
\begin{equation}
\xi_{\mathcal{E}}=(\mathcal{E}\otimes I)(\Phi_{+}).
\end{equation}
By teleporting over this state can we simulate channel $\mathcal{E}$?

\begin{figure}[ptb]
\vspace{0.2cm}
\centering\includegraphics[width=0.41\textwidth]{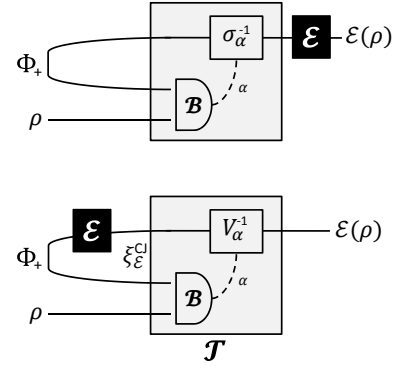}
\vspace{-0.2cm}\caption{(Top) Consider a qubit teleportation
protocol where an input state $\rho$ is teleported to the input of
a quantum channel $\mathcal{E}$. This is achieved by applying a
Bell detection $\mathcal{B}$ to the input $\rho$ and a
maximally-entangled state $\Phi_{+}$, followed by the classical
communication of the outcome $\alpha$ which triggers a conditional
Pauli correction $\sigma_{\alpha}^{-1}$. (Bottom) Assume that
$\mathcal{E}$ is teleportation covariant as in
Eq.~(\ref{telecov}). The Pauli corrections can be pushed at the
output of the channel where they become generally-different
unitary corrections $V_{\alpha}^{-1}$ (depending on the channel
these may again be Pauli operators or not). Now the application of
the channel
$\mathcal{E}$ on $\Phi_{+}$ creates the Choi matrix $\xi_{\mathcal{E}%
}^{\text{CJ}}=(\mathcal{E}\otimes I)(\Phi_{+})$ as a resource state for the
next teleportation protocol $\mathcal{T}$. As a result the channel
$\mathcal{E}$ is simulated by teleporting over its Choi matrix as in
Eq.~(\ref{topp}). We also say that a teleportation-covariant channel
$\mathcal{E}$ is a Choi-stretchable channel~\cite{PLOB}.}%
\label{Fig:telep}%
\end{figure}

The answer is \textit{yes} for so-called teleportation-covariant
channels~\cite{MHthesis,Leung,PLOB}. By definition a quantum channel
$\mathcal{E}$ is teleportation-covariant if, for any random teleportation
unitary $U$ (corresponding to a Pauli operator in the qubit case), we may
write
\begin{equation}
\mathcal{E}(U\rho U^{\dagger})=V\mathcal{E}(\rho)V^{\dagger}, \label{telecov}%
\end{equation}
for some other unitary $V$~\cite{PLOB}. This property is a sufficient
condition to ensure that the channel $\mathcal{E}$ can be simulated by
teleporting over its Choi matrix or Choi-Jamiolkowski (CJ) state
$\xi_{\mathcal{E}}^{\text{CJ}}$ (this is also known as teleportation-simulable
or Choi-stretchable channel~\cite{PLOB}). In other words, we may write the
simulation~\cite{PLOB,nostro}
\begin{equation}
\mathcal{E}(\rho)=\mathcal{T}(\rho\otimes\xi_{\mathcal{E}}^{\text{CJ}}),
\label{topp}%
\end{equation}
where $\mathcal{T}$ is teleportation. See Fig.~\ref{Fig:telep} for a visual
proof of Eq.~(\ref{topp}). This is clearly a powerful design but only holds
for the teleportation-covariant subset of programmable channels.

In the setting of quantum parameter estimation, we are interested in joint
teleportation-covariance, where a parametrised quantum channel $\mathcal{E}%
_{\theta}$ satisfies Eq.~(\ref{telecov}) with a $\theta$-independent set of
output unitaries, i.e.,~\cite{nostro}
\begin{equation}
\mathcal{E}_{\theta}(U\rho U^{\dagger})=V\mathcal{E}_{\theta}(\rho)V^{\dagger
},~~\text{for any }\theta\text{.} \label{coTELE}%
\end{equation}
This is exactly the situation when $\theta$ is a noise parameter, i.e., a
parameter that can be uniquely associated to an environment dilating the
quantum channel.

If Eq.~(\ref{coTELE}) holds, then we can write $\mathcal{E}_{\theta}%
(\rho)=\mathcal{T}(\rho\otimes\xi_{\mathcal{E}_{\theta}}^{\text{CJ}})$ and
repeat the stretching of a quantum comb as before. In this way, we may
decompose the output state of an adaptive parameter estimation protocol
as~\cite{nostro}%
\begin{equation}
\rho_{\theta}^{n}=\Lambda\left[  (\xi_{\mathcal{E}_{\theta}}^{\text{CJ}%
})^{\otimes n}\right]  ,
\end{equation}
for some quantum channel $\Lambda$. As a result, we get%
\begin{equation}
\mathrm{QFI}(\rho_{\theta}^{n})\leq n\mathrm{QFI}(\xi_{\mathcal{E}_{\theta}%
}^{\text{CJ}}). \label{ChoiQFI}%
\end{equation}
This means that the estimation of a noise parameter of a
teleportation-covariant channel is limited to the SQL with a pre-factor given
by its Choi matrix, i.e., the QCRB reads~\cite{nostro}
\begin{equation}
\delta\theta^{2}\geq\lbrack n\mathrm{QFI}(\xi_{\mathcal{E}_{\theta}%
}^{\text{CJ}})]^{-1}. \label{QCRBmmm}%
\end{equation}

The teleportation simulation not only allows us to compute explicitly the
upper bound, but also yields a matching lower bound. As a matter of fact, an
optimal strategy that saturates the bound employs a block (assisted)
estimation protocol where the maximally-entangled state $\Phi_{+}$ is used at
the input of the channel in an identical and independent way. This strategy
provides a QFI exactly equal to $n\mathrm{QFI}(\xi_{\mathcal{E}_{\theta}%
}^{\text{CJ}})$. As a result, the QCRB in Eq.~(\ref{QCRBmmm}) is
asymptotically achievable for large $n$.

Let us compute the QCRB for specific examples. It is known that erasure,
dephasing and depolarizing channels are teleportation-covariant~\cite{PLOB}.
More precisely, these channels satisfy the condition of joint teleportation
covariance of Eq.~(\ref{coTELE}) with $\theta$ being their channel-defining
probability parameter $p$. Recall that an erasure channel is represented
by~\cite{NiCh}
\begin{equation}
\mathcal{E}_{p}^{\text{erase}}(\rho)=(1-p)\rho+p\left\vert e\right\rangle
\left\langle e\right\vert ,
\end{equation}
where $\left\vert e\right\rangle $ is an orthogonal erasure state and $p$ is
the erasure probability. A dephasing channel is defined as~\cite{NiCh}%
\begin{equation}
\mathcal{E}_{p}^{\text{phase}}(\rho)=(1-p)\rho+pZ\rho Z^{\dagger},
\end{equation}
where $p$\ is the probability of phase flip. Finally, a depolarizing channel
with probability $p$ is defined as~\cite{NiCh}
\begin{equation}
\mathcal{E}_{p}^{\text{depol}}(\rho)=(1-p)\rho+p\pi,
\end{equation}
where $\pi$ is the maximally-mixed state.

For each family of these channels $\mathcal{E}_{p}$ (i.e., erasure, dephasing
or depolarizing),\ we compute the Choi matrix $\xi_{\mathcal{E}_{p}%
}^{\text{CJ}}$ and the associated QFI, finding $\mathrm{QFI}(\xi
_{\mathcal{E}_{p}}^{\text{CJ}})=[p(1-p)]^{-1}$ for each of the families. Then,
using Eq.~(\ref{QCRBmmm}), we find that the adaptive estimation of $p$ is
bounded by the following asymptotically-achievable QCRB~\cite{nostro}%
\begin{equation}
\delta p^{2}\geq p(1-p)n^{-1}.
\end{equation}

\section{Extension to continuous variables\label{Sec4}}

\subsection{Teleportation simulation of bosonic channels}

In this section we consider bosonic channels and their teleportation
simulation. We start by reviewing the teleportation of bosonic states
\textit{\`{a} la} Vaidman~\cite{Vaid} and then \textit{\`{a} la} Braunstein
and Kimble~\cite{BK}. We then discuss how the latter protocol can be modified
to simulate bosonic channels and, in particular, bosonic Gaussian
channels~\cite{Cirac,Niset,PLOB,nostro}. The optimal simulation of bosonic
channels is asymptotic and requires a careful treatment of the simulation
error by introducing a suitable energy-bounded diamond norm. We therefore
follow the formalism developed in Refs.~\cite{PLOB,nostro} which rigorously
accounts for these aspects (see also Ref.~\cite{TQCreview}).

Consider a bosonic mode with quadrature operators $\hat{q}$, $\hat{p}$
satisfying the commutation relation $[\hat{q},\hat{p}]=i$ (we put $\hbar=1$).
A bosonic channel is a CPTP map between an input and an output mode. Vaidman's
teleportation protocol~\cite{Vaid} considers an ideal (infinite-energy) EPR
state $\Phi_{\text{EPR}}$ of modes $A$ (Alice) and $B$ (Bob). An input mode
$a$, prepared in some finite-energy state $\rho$, is then mixed in a balanced
beam-splitter with mode $A$ and the two output modes \textquotedblleft$\pm
$\textquotedblright\ are homodyned with outcomes $q_{-}$ and $p_{+}$. This
detection realises the ideal continuous-variable Bell detection $\mathcal{B}%
$\ (which projects on displaced EPR states). The complex variable
$\alpha=q_{-}+ip_{+}$ is then sent to Bob, who applies a
displacement~\cite{RMPwee} $D(-\alpha)$ on his mode $B$, thus retrieving the
input state $\rho$.

The Braunstein-Kimble protocol~\cite{BK} removes the singularities from the
previous description, therefore allowing for a realistic and practical
implementation of the idea~\cite{teleEXP}. The main point is to use a two-mode
squeezed vacuum (TMSV) state $\Phi_{\mu}$ as resource for teleportation. This
is a two-mode Gaussian state~\cite{RMPwee} with zero mean and $\mu$-dependent
covariance matrix. Its parameter $\mu$ quantifies both the amount of two-mode
squeezing (or entanglement) between modes $A$ and $B$, and the variance of the
thermal noise in each individual mode. The ideal EPR\ state can be defined by
taking the limit for infinite squeezing, i.e., we may define the asymptotic
state $\Phi_{\text{EPR}}:=\lim_{\mu}\Phi_{\mu}$ in terms of a diverging
sequence of TMSV states. Similarly, the same relaxation can be done for the
Bell detection. One may consider a sequence of Gaussian POVMs~\cite{RMPwee}
$\mathcal{B}_{\mu}$ which are (quasi-)projections on displaced TMSV states
$\Phi_{\mu,\alpha}:=D(\alpha)\Phi_{\mu}D(-\alpha)$. The ideal case is obtained
by taking the limit of $\mu\rightarrow\infty$, i.e., the ideal Bell detection
is formally defined as $\mathcal{B}:=\lim_{\mu}$ $\mathcal{B}_{\mu}$.

It is clear that, using a realistic Braunstein-Kimble protocol with finite
squeezing $\mu$ (for both resource and Bell detection), we cannot achieve
perfect teleportation fidelity. However, we may asymptotically approximate
perfect teleportation for large values of $\mu$ for any energy-bounded
alphabet at the input~\cite{BK,twist}. In other words, consider the compact
set of energy-constrained single-mode bosonic states $\mathcal{D}_{N}%
^{1}:=\{\rho:\mathrm{Tr}(\rho\hat{N})\leq N\}$ where $\hat{N}$ is the photon
number operator. For any input $\rho\in\mathcal{D}_{N}^{1}$, we write the
output of the Braunstein-Kimble $\mu$-protocol $\mathcal{T}_{\mu}$ as
$\rho_{\mu}:=\mathcal{E}_{\mu}^{\text{BK}}(\rho)$, where $\mathcal{E}_{\mu
}^{\text{BK}}$ is an associated teleportation channel. In the limit of large
$\mu$, one has
\begin{equation}
||\rho_{\mu}-\rho||\overset{\mu}{\rightarrow}0, \label{pointwise}%
\end{equation}
for any finite $N$. This result may be extended to the presence of an
ancillary system and mapped into a corresponding convergence in energy-bounded
diamond distance.

Let us define the set of energy-constrained bipartite states
\begin{equation}
\mathcal{D}_{N}:=\{\rho_{ra}~:~\mathrm{Tr}(\hat{N}_{ra}\rho_{ra})\leq N\},
\label{finiteALFA}%
\end{equation}
where $r$ is an arbitrary ancillary multi-mode system and $\hat{N}_{ra}$ is
the total number operator. One can check that $\mathcal{D}_{N}$ is a compact
set~\cite{HolevoCOMPACT}. Then, for two bosonic channels, $\mathcal{E}_{1}$
and $\mathcal{E}_{2}$, we may define the energy-bounded diamond distance
as~\cite{PLOB,nostro}%
\begin{equation}
\left\Vert \mathcal{E}_{1}-\mathcal{E}_{2}\right\Vert _{\diamond N}%
:=\sup_{\rho_{ra}\in\mathcal{D}_{N}}\Vert\mathcal{I}_{r}\otimes\mathcal{E}%
_{1}(\rho_{ra})-\mathcal{I}_{r}\otimes\mathcal{E}_{2}(\rho_{ra})\Vert~.
\label{defBBBB}%
\end{equation}
(See Ref.~\cite{Shirokov} for a slightly different definition of
energy-constrained diamond norm). For any energy constraint $N$, consider the
distance between the Braunstein-Kimble channel $\mathcal{E}_{\mu}^{\text{BK}}$
and the identity channel $\mathcal{I}$ associated with perfect teleportation
(\textit{\`{a} la} Vaidman). From the point-wise trace-norm limit in
Eq.~(\ref{pointwise}) and the compactness of $\mathcal{D}_{N}$, we derive the
vanishing simulation error
\begin{equation}
\delta(\mu,N):=\left\Vert \mathcal{E}_{\mu}^{\text{BK}}-\mathcal{I}\right\Vert
_{\diamond N}\overset{\mu}{\rightarrow}0,~~\text{for any finite }N\text{.}%
\end{equation}
Here it is important to remark that the latter convergence to zero is not
guaranteed if we consider unconstrained alphabets, i.e., we remove $N<+\infty
$. It is in fact easy to construct a sequence of input states with diverging
energy $N$ such that the joint limit of the error simulation $\delta(\mu,N)$
in $N$ and $\mu$ is not defined. For this counter-example see discussions in
Ref.~\cite{TQCreview}.

Consider now a teleportation-covariant bosonic channel. This means that the
channel must satisfy the property~\cite{PLOB}%
\begin{equation}
\mathcal{E}[D(\alpha)\rho D(-\alpha)]=D(\tilde{\alpha})\mathcal{E}%
(\rho)D(-\tilde{\alpha})
\end{equation}
where the output amplitudes $\tilde{\alpha}$ are functions of the input ones
$\alpha$. This is certainly the case for single-mode Gaussian
channels~\cite{RMPwee}. Because of this property, we may write the
continuous-variable version of the simulation in Eq.~(\ref{topp}). In fact, by
simulating $\mathcal{E}$ with a finite-squeezing Braunstein-Kimble protocol
$\mathcal{T}_{\mu}$, we generate the approximated channel
\begin{equation}
\mathcal{E}_{\mu}(\rho)=\mathcal{T}_{\mu}(\rho\otimes\xi_{\mathcal{E}}^{\mu}),
\label{teleFINNN}%
\end{equation}
where $\mathcal{T}_{\mu}$ is based on a finite-squeezing Bell detection
$\mathcal{B}_{\mu}$ and $\xi_{\mathcal{E}}^{\mu}$ is generated by a TMSV state
$\Phi_{\mu}$ as%
\begin{equation}
\xi_{\mathcal{E}}^{\mu}:=(\mathcal{E}\otimes I)(\Phi_{\mu}).
\end{equation}
The latter defines the asymptotic Choi matrix in the limit $\xi_{\mathcal{E}%
}^{\text{CJ}}:=\lim_{\mu}\xi_{\mathcal{E}}^{\mu}$. Note that we may write the
composition $\mathcal{E}_{\mu}=\mathcal{E}\circ\mathcal{E}_{\mu}^{\text{BK}}%
$.\ Therefore, for any bounded alphabet with energy $N$, we have the channel
simulation error~\cite{PLOB,nostro}%
\begin{equation}
\left\Vert \mathcal{E}_{\mu}-\mathcal{E}\right\Vert _{\diamond N}%
\leq\left\Vert \mathcal{E}_{\mu}^{\text{BK}}-\mathcal{I}\right\Vert _{\diamond
N}:=\delta(\mu,N)~.
\end{equation}

\subsection{Teleportation stretching of a comb in continuous variables}

Assume now that the quantum channel $\mathcal{E}$ fills $n$ slots of a quantum
comb with output $\rho^{n}$. Then, assume to replace $\mathcal{E}$ with its
imperfect simulation $\mathcal{E}_{\mu}$ so that the output is $\rho_{\mu}%
^{n}$. We may bound the simulation error on the output state $||\rho_{\mu}%
^{n}-\rho^{n}||$ in terms of the channel simulation error. In fact, by
adopting a peeling argument~\cite{PLOB,nostro} based on basic properties of
the trace distance (i.e., its monotonicity under CPTP maps and the triangle
inequality), we may write~\cite{PLOB}%
\begin{equation}
||\rho_{\mu}^{n}-\rho^{n}||\leq n\left\Vert \mathcal{E}_{\mu}-\mathcal{E}%
\right\Vert _{\diamond N}\leq n\delta(\mu,N)~. \label{eq1rr}%
\end{equation}
Then, we also observe that we may stretch the approximated channel
$\mathcal{E}_{\mu}$ by using the teleportation simulation of
Eq.~(\ref{teleFINNN}). Therefore, for the simulated output we may write the
decomposition~\cite{PLOB}
\begin{equation}
\rho_{\mu}^{n}=\Lambda_{\mu}\left[  (\xi_{\mathcal{E}}^{\mu})^{\otimes
n}\right]  , \label{eq2rr}%
\end{equation}
where $\Lambda_{\mu}$ is a global quantum channel associated with the quantum
comb and also the teleportation protocol $\mathcal{T}_{\mu}$.

Thus, combining Eqs.~(\ref{eq1rr}) and~(\ref{eq2rr}), we may
write~\cite{PLOB}
\begin{equation}
\left\Vert \rho^{n}-\Lambda_{\mu}(\xi_{\mathcal{E}}^{\mu\otimes n})\right\Vert
\leq n\delta(\mu,N)~, \label{correctS}%
\end{equation}
which goes to zero for large $\mu$ and finite $N$ (and $n$). The latter
Eq.~(\ref{correctS}) represents the rigorous stretching of an adaptive
protocol (quantum comb) performed over a teleportation-covariant bosonic channel.

As discussed in Ref.~\cite{TQCreview} in relation to the use of channel
simulation in quantum/private communications, other approaches that neglect
the energy constraint on the input alphabet and do not explicitly describe the
propagation of the simulation error from the channels to the output state may
be affected by technical issues and divergences in the results.

\subsection{Teleportation stretching of adaptive metrology in continuous
variables}

To apply the methodology to adaptive parameter estimation, we need joint
teleportation covariance for the family of channels $\mathcal{E}_{\theta}$
spanned by varying the parameter $\theta$. If this is the case, then we may
repeat the previous procedure and decompose the output state $\rho_{\theta
}^{n}$ by using~\cite{nostro}
\begin{equation}
\left\Vert \rho_{\theta}^{n}-\Lambda_{\mu}(\xi_{\mathcal{E}_{\theta}}%
^{\mu\otimes n})\right\Vert \leq n\delta(\mu,N)~,
\end{equation}
for any $\theta$, finite number of uses $n$ and finite energy $N$. To evaluate
the QFI of $\rho_{\theta}^{n}$, we now exploit the connection with the Bures
distance $d_{\text{B}}$ and the trace distance $D$. In fact, we may write%
\begin{equation}
\mathrm{QFI}(\rho_{\theta}^{n})=\frac{4d_{\text{B}}^{2}(\rho_{\theta}^{n}%
,\rho_{\theta+d\theta}^{n})}{d\theta^{2}},
\end{equation}
where
\begin{align}
d_{\text{B}}(\rho,\sigma)  &  :=\sqrt{2[1-F(\rho,\sigma)]}\nonumber\\
&  \leq\sqrt{2D(\rho,\sigma)}=\sqrt{||\rho-\sigma||}.
\end{align}

Using the triangle inequality for the Bures distance and properties of the
fidelity (monotonicity under CPTP\ maps and multiplicativity over tensor
products), we may write~\cite{nostro}%
\begin{equation}
d_{\text{B}}(\rho_{\theta}^{n},\rho_{\theta+d\theta}^{n})\leq\sqrt
{2[1-(F_{\theta}^{\mu})^{n}]}+2\sqrt{n\delta(\mu,N)},
\end{equation}
where $F_{\theta}^{\mu}:=F(\xi_{\mathcal{E}_{\theta}}^{\mu},\xi_{\mathcal{E}%
_{\theta+d\theta}}^{\mu})$. For any finite $n$ and $N$, we may take the limit
for large $\mu$ and write%
\begin{equation}
d_{\text{B}}(\rho_{\theta}^{n},\rho_{\theta+d\theta}^{n})\leq\lim_{\mu}%
\sqrt{2[1-(F_{\theta}^{\mu})^{n}]}=\sqrt{2[1-(F_{\theta}^{\infty})^{n}]}~,
\end{equation}
where $F_{\theta}^{\infty}:=\lim_{\mu}F_{\theta}^{\mu}$. In other words, we
have%
\begin{equation}
\mathrm{QFI}(\rho_{\theta}^{n})\leq\frac{8[1-(F_{\theta}^{\infty})^{n}%
]}{d\theta^{2}}.
\end{equation}
It is easy to check~\cite{nostro} that the upper bound is additive, so that
\begin{equation}
\mathrm{QFI}(\rho_{\theta}^{n})\leq n\frac{8[1-F_{\theta}^{\infty}]}%
{d\theta^{2}}:=n\mathrm{QFI}_{\theta}^{\infty}. \label{QFIcv}%
\end{equation}

It is important to note that the upper bound does not depend on the specifics
of the adaptive protocol and also on energy constraint $N$. Therefore, the
bound is valid for all possible adaptive protocols, both constrained and
unconstrained (i.e., we can safely remove the energy constraint at the end of
the calculations). Also notice that the upper bound is asymptotically
achievable by an unconstrained block (assisted) protocol, where $n$ TMSV
states $\Phi_{\mu}$ are used to probe the channel, so that one collects the
output product state $\xi_{\mathcal{E}_{\theta}}^{\mu\otimes n}$. By making an
optimal measurement, we achieve
\begin{equation}
\mathrm{QFI}(\xi_{\mathcal{E}_{\theta}}^{\mu\otimes n})=n\frac{8[1-F_{\theta
}^{\mu}]}{d\theta^{2}},
\end{equation}
whose limit for large $\mu$ coincides with the upper bound in Eq.~(\ref{QFIcv}%
). Because, this protocol uses independent probing states, the QCRB is
achievable for large $n$.

In conclusion, Eq.~(\ref{QFIcv}) is indeed the ultimate QFI achievable with
adaptive estimation protocols. Thus, we may say that the optimal adaptive
estimation of a noise parameter $\theta$ encoded in a teleportation-covariant
bosonic channel $\mathcal{E}_{\theta}$ (so that the family is jointly
tele-covariant)\ is limited to the SQL. In fact, it satisfies the
asymptotically achievable QCRB~\cite{nostro}%
\begin{equation}
\delta\theta^{2}\geq(n\mathrm{QFI}_{\theta}^{\infty})^{-1}~, \label{QCRBcvv}%
\end{equation}
where $\mathrm{QFI}_{\theta}^{\infty}$ is related to the asymptotic Choi
matrix of the channel $\xi_{\mathcal{E}_{\theta}}^{\text{CJ}}$ according to
the limit in Eq.~(\ref{QFIcv}).

\subsection{Results for bosonic Gaussian channels}

Consider a single bosonic mode with quadrature operators $\mathbf{\hat{x}%
}=(\hat{q},\hat{p})^{T}$. A Gaussian state is completely characterised by its
mean value $\mathbf{\bar{x}}$ and covariance matrix (CM) $\mathbf{V}%
$~\cite{RMPwee}. A single-mode Gaussian channel transforms these statistical
moments as follows
\begin{equation}
\mathbf{\bar{x}}\rightarrow\mathbf{T\bar{x}}+\mathbf{d},~~\mathbf{V}%
\rightarrow\mathbf{TVT}^{T}+\mathbf{N,} \label{Act}%
\end{equation}
where $\mathbf{d}$ is a displacement vector, $\mathbf{T}$ and $\mathbf{N}%
=\mathbf{N}^{T}$ are $2\times2$ real matrices satisfying the conditions
$\mathbf{N}=\mathbf{N}^{T}\geq0$ and $\det\mathbf{N}\geq(\det\mathbf{T}%
-1)^{2}/4$~\cite{Caruso,RMPwee}. Phase-insensitive Gaussian channels have
diagonal matrices
\begin{equation}
\mathbf{T}=\sqrt{\eta}\,\mathbf{I},~~\mathbf{N}=\nu\mathbf{I} \label{1mode}%
\end{equation}
where $\eta\in\mathbb{R}$ is a transmissivity parameter (loss or
amplification), while $\nu\geq0$\ represents noise~\cite{RMPwee}. Typically,
they also have $\mathbf{d}=\mathbf{0}$, i.e., they do not add displacements to
the input.

One of the most important is the thermal-loss channel $\mathcal{E}_{\eta
,\bar{n}}^{\text{loss}}$, which is defined by transmissivity $\eta\in
\lbrack0,1]$ and noise $\nu=(1-\eta)(\bar{n}+1/2)$ with thermal number
$\bar{n}$. This channel can be realised by a beam-splitter (of transmissivity
$\eta$) mixing the input with an environmental thermal mode with $\bar{n}$
mean number of photons. It is clearly teleportation-covariant. More strongly,
it is jointly teleportation-covariant in the thermal number $\bar{n}$.
Therefore, consider the adaptive estimation of parameter $\bar{n}>0$ (e.g.,
this can be related to a measurement of temperature). By using
Eq.~(\ref{QCRBcvv}) one computes~\cite{nostro} $\mathrm{QFI}_{\bar{n}}%
^{\infty}=[\bar{n}(\bar{n}+1)]^{-1}$ and therefore the QCRB%
\begin{equation}
\delta\bar{n}^{2}\geq\frac{\bar{n}(\bar{n}+1)}{n}~. \label{bbb}%
\end{equation}
We see that the QCRB does not depend on the loss parameter $\eta$, as long as
it is less than $1$. This implies that, for any $\eta<1$, we achieve the same
accuracy as we would get in a direct measurement of the environment ($\eta=0$).

Consider now a noisy quantum amplifier $\mathcal{E}_{\eta,\bar{n}}%
^{\text{amp}}$ which is defined by a gain $\eta>1$ and noise $\nu
=(\eta-1)(\bar{n}+1/2)$ with thermal number $\bar{n}$. This is teleportation
covariant and jointly tele-covariant in the parameter $\bar{n}$. For the
adaptive estimation of $\bar{n}>0$, one gets~\cite{nostro} the same QCRB of
Eq.~(\ref{bbb}). Finally, consider an additive-noise Gaussian channel
$\mathcal{E}_{\nu}^{\text{add}}$ which is defined by $\eta=1$ and $\nu\geq0$.
This is joint teleportation covariant in the added noise $\nu$, whose optimal
adaptive estimation is bounded by~\cite{nostro} $\mathrm{QFI}_{\nu}^{\infty
}=\nu^{-2}$ and therefore the QCRB
\begin{equation}
\delta\nu^{2}\geq\nu^{2}/n~. \label{addMM}%
\end{equation}

\section{Sub-optimal simulation of bosonic Gaussian channels\label{Sec5}}

Here we present an alternative simulation for single-mode bosonic Gaussian
channels which does not need to consider the limit of an asymptotic Choi
matrix (but still requires the limit of an ideal Bell detection). Consider a
two-mode Gaussian state with zero mean and generic CM
\begin{equation}
\mathbf{V}_{AB}=\left(
\begin{array}
[c]{cc}%
\mathbf{A} & \mathbf{C}\\
\mathbf{C}^{T} & \mathbf{B}%
\end{array}
\right)  \,. \label{stan}%
\end{equation}
By teleporting over this Gaussian resource using a Braunstein-Kimble protocol
with gain $g$ we obtain a Gaussian teleportation channel such
that~\cite{finite} $\mathbf{\bar{x}}\rightarrow g\mathbf{\bar{x}}$ and%
\begin{equation}
\mathbf{V}\rightarrow g^{2}\mathbf{V}+g^{2}\mathbf{Z}\mathbf{A}\mathbf{Z}%
+\mathbf{B}-g(\mathbf{Z}\mathbf{C}+\mathbf{C}^{T}\mathbf{Z})~,
\end{equation}
where $\mathbf{Z}=\mathrm{diag}(1,-1)$. Therefore, a phase-insensitive
Gaussian channel $\mathcal{E}_{\eta,\nu}$ with parameters $\eta$ and $\nu$
[see Eqs.~(\ref{Act}) and~(\ref{1mode}) with $\mathbf{d}=\mathbf{0}$] can be
simulated by using the gain $g=\sqrt{\eta}$ and using a CM $\mathbf{V}_{AB}$
with the choice
\begin{equation}
\mathbf{A}=a\mathbf{I},~\mathbf{B}=b\mathbf{I},~\mathbf{C}=c\mathbf{Z,}%
\end{equation}
so that $\nu=ag^{2}-2cg+b$~\cite{finite}.

We are interested in finding a finite-energy resource state $\sigma_{\nu}$
that can simulate a phase-insensitive Gaussian channel $\mathcal{E}_{\eta,\nu
}$ according to
\begin{equation}
\mathcal{E}_{\eta,\nu}(\rho)=\mathcal{T}_{\eta}(\rho\otimes\sigma_{\nu})~,
\label{Finn}%
\end{equation}
where $\mathcal{T}_{\eta}$ is the Braunstein-Kimble protocol with ideal Bell
detection and gain $g=\sqrt{\eta}$. More precisely, we may write
$\mathcal{T}_{\eta}=\lim_{\mu}\mathcal{T}_{\mu}^{\eta}$, where $\mathcal{T}%
_{\mu}^{\eta}$ is the Braunstein-Kimble $\mu$-protocol with gain $g=\sqrt
{\eta}$. A possible choice for $\sigma_{\nu}$ is a Gaussian state with zero
mean and CM
\begin{equation}
\mathbf{V}(\sigma_{\nu})=\left(
\begin{array}
[c]{cc}%
a\mathbf{I} & c\mathbf{Z}\\
c\mathbf{Z} & b\mathbf{I}%
\end{array}
\right)  , \label{resState}%
\end{equation}
with the following elements
\begin{equation}
a=\frac{1}{2}\cosh{2r},~b=\frac{|1-\eta|}{2}+\frac{\eta}{2}\cosh{2r}%
,~c=\frac{\sqrt{\eta}}{2}\sinh{2r}~, \label{lakk}%
\end{equation}
where%
\begin{equation}
r=-\frac{1}{2}\ln{\left[  \frac{2\nu-|1-\eta|}{2\eta}\right] } ~.
\label{lastGG}%
\end{equation}

It is worth remarking that there exist many finite-energy resource states that
can simulate a given channel. A different family of resource states has been
obtained in Ref.~\cite{finite} to characterise the teleportation fidelity.
This family of resource states has also been exploited in quantum
communication~\cite{finiteStre} to derive weak converse upper bounds for the
secret key capacity of phase-insensitive Gaussian channels. These bounds
closely approximate the ideal and tightest bounds obtained for infinite
energy~\cite{PLOB}. In what follows we use the sub-optimal simulation of
Eq.~(\ref{Finn}) with the finite-energy resource state specified by
Eqs.~(\ref{resState})-(\ref{lastGG}). It is the first time that this
finite-resource approach is used in quantum metrology.

Note that the form of the simulation in Eq.~(\ref{Finn}) is such that the
noise parameter $\nu$ only appears in the resource state $\sigma_{\nu}$ or, in
other words, the teleportation LOCC $\mathcal{T}_{\eta}$ does not depend on
$\nu$. For this reason, the family of channels $\mathcal{E}_{\eta,\nu}$ with
fixed $\eta$ but varying $\nu$ is a family of jointly teleportation-simulable
channels (which is a condition implied by the joint teleportation covariance).
As a result, the adaptive estimation of the parameter $\nu$ can be completely
simplified, so that the $n$-use output state of a comb reads $\rho_{\nu}%
^{n}=\Lambda_{\eta}(\sigma_{\nu}^{\otimes n})$ for some global quantum channel
$\Lambda_{\eta}$ which is independent from the unknown parameter $\nu$. As a
consequence, we may simplify the QFI of the output state $\rho_{\nu}^{n}$ and
write the following QCRB for the adaptive estimation of $\nu$
\begin{equation}
\delta\nu^{2}\geq\frac{1}{n\mathrm{QFI}(\sigma_{\nu})}~. \label{finQCRB}%
\end{equation}

As an example consider the additive-noise Gaussian channel $\mathcal{E}_{\nu
}^{\text{add}}$. This channel can be simulated by exploiting a resource state
$\sigma_{\nu}$ whose CM\ is given by Eq.~(\ref{resState})-(\ref{lastGG}) with
$\eta=1$ (see also Refs.~\cite{Ban,Andy}). We may then compute the QFI from
the quantum fidelity~\cite{Banchi}, and find the QCRB $\delta v^{2}\geq
v^{2}/n$. Note that this exactly coincides with the tight achievable bound of
Eq.~(\ref{addMM}) which is obtained by simulating the channel via its
asymptotic Choi matrix.

\begin{figure}[h] \centering\vspace{0.0cm}
\includegraphics[width=0.28\textwidth]{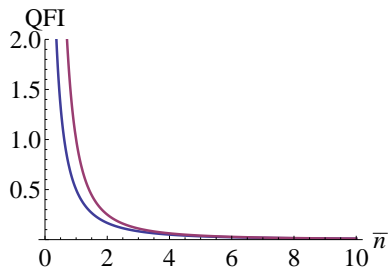}
\vspace{-0.0cm} \caption{Quantum Fisher information
$\mathrm{QFI}(\sigma _{\bar{n}})$ associated with the adaptive
estimation of the thermal number $\bar{n}$ of a thermal-loss
channel $\mathcal{E}_{\eta,\bar{n}}$. Assuming the sub-optimal
simulation we find $\mathrm{QFI}(\sigma_{\bar{n}})=\bar{n}^{-2}$
(upper red line). Compare this with
$\mathrm{QFI}_{\bar{n}}^{\infty}=[\bar {n}(\bar{n}+1)]^{-1}$ which
is computed using the asymptotic simulation (lower
blue line).}%
\label{Fig:finiteQFI}%
\end{figure}

Consider now the adaptive estimation of the thermal number $\bar{n}$ of a
thermal-loss channel $\mathcal{E}_{\eta,\bar{n}}^{\text{loss}}$ assuming the
sub-optimal simulation. Putting $\nu=(1-\eta)(\bar{n}+1/2)$ in
Eq.~(\ref{finQCRB}) we compute the QCRB for $\delta\bar{n}^{2}$. We do not
find the tight achievable bound of Eq.~(\ref{bbb}) but a larger bound given
by
\begin{equation}
\delta\bar{n}^{2}\geq\bar{n}^{2}/n\,.\label{eq60}%
\end{equation}
For comparison, in Fig.~\ref{Fig:finiteQFI} we plot the QFI for
the asymptotic and finite-energy resource state. It is a open
problem to find a finite-energy resource that can match the
asymptotic bound. Finally, one may easily check that
Eq.~(\ref{eq60}) also holds for a noisy amplifier
$\mathcal{E}_{\eta ,\bar{n}}^{\text{amp}}$ assuming its
sub-optimal simulation. 

\section{Conclusions\label{Sec6}}

Channel simulation is a powerful tool for completely simplifying protocols of
adaptive parameter estimation, for instance represented as a quantum comb.
This technique allows one to compute the ultimate precision in estimating
noise parameters that are encoded in discrete- or continuous-variable
channels. The tool easily applies to any programmable channel whose unknown
parameter is encoded in its program (environmental) state. One can then reduce
an adaptive protocol and show that the QCRB is limited to the SQL.

When a programmable channel is teleportation-covariant (such as an erasure, a
Pauli or a Gaussian channel), we can exploit a precise design for its
simulation which is based on a simple modification of the teleportation
protocol. In this way, we may show that the QCRB is limited to the SQL with
the QFI being computed on the Choi matrix of the channel (in an asymptotic
fashion for bosonic Gaussian channels). Furthermore, the QCRB is shown to be
achievable by a block (i.e., non-adaptive) protocol based on entanglement-assistance.

As a consequence of the previous results, a quantum channel able to beat the
SQL and potentially reach the Heisenberg scaling must be necessarily
non-programmable in the sense discussed in this review, i.e., it cannot be
perfectly simulated by means of a single-copy program state. A potential
approach to cover this type of channel is therefore considering an extended
definition of multi-programmability where the simulation is achieved by using
a multi-copy resource state.

In conclusion, we have reviewed the state-of-the-art in the theory of channel
simulation within the context of quantum parameter estimation. The reader
interested in similar applications in quantum channel discrimination may
consult Ref.~\cite{nostro} and a forthcoming review paper~\cite{NatPHOTONICs}.
The reader interested in applications to quantum and private communications
(e.g., for establishing two-way capacities) may consult Ref.~\cite{TQCreview}
and also one of the founding papers~\cite{PLOB}.


\textbf{Acknowledgments.~}This work has been supported by the Innovation Fund
Denmark (Qubiz project), the European Union (MSCA-IF-2016), and the EPSRC via
the `UK Quantum Communications Hub' (EP/M013472/1). S. P. thanks R.
Demkowicz-Dobrza\'{n}ski, S. Lloyd, L. Maccone, J. Kolodynski, D. Braun, and
G. Adesso for feedback on these topics.

\end{document}